\documentclass[aps,pra,twocolumn]{revtex4-1}
\usepackage{amssymb}
\usepackage{graphicx}
\usepackage{color}
\usepackage{epstopdf}
\usepackage{amsmath}
\usepackage{bm}
\usepackage{cleveref}
\usepackage{tcolorbox}
\usepackage[export]{adjustbox}

\newcommand{\reffig}[1]{Fig.~\ref{#1}}

\begin{document}

\title{Phases of supersolids in confined dipolar Bose-Einstein condensates}

\author{Yong-Chang Zhang$^{1,2}$}
    \author{Thomas Pohl$^{1}$}
\author{Fabian Maucher$^{1,3}$}
\affiliation{$^1$Center for Complex Quantum Systems, Department of Physics and Astronomy, Aarhus University, DK-8000 Aarhus C, Denmark\\
$^2$MOE Key Laboratory for Nonequilibrium Synthesis and Modulation of Condensed Matter, Shaanxi Key Laboratory of Quantum Information and Quantum Optoelectronic Devices, School of Physics, Xi'an Jiaotong University, Xi'an 710049, People's Republic of China\\
$^3$Departament de Física, Universitat de les Illes Balears \& IAC-3, Campus UIB, E-07122 Palma de Mallorca, Spain}

\begin{abstract}
Dipolar Bose-Einstein condensates represent a powerful platform for the exploration of quantum many-body phenomena arising from long-range interactions. A series of recent experiments has demonstrated the formation of supersolid states of matter. Subsequent theoretical works have shown that quantum fluctuations can affect the underlying phase transition and may lead to the emergence of supersolids with various lattice structures in dipolar condensates. In this work we explore the signatures of such different geometries in confined finite condensates. In addition to previously found triangular lattices, our analysis reveals a rich spectrum of states, from honeycomb patterns 
to striped supersolids. By optimizing relevant parameters we show that transitions between distinct supersolids should be observable in current experiments. 
\end{abstract}
\maketitle

\section{Introduction}
Long-range particle interactions in Bose-Einstein condensates (BECs) can promote a range of 
intriguing physical phenomena, such as the formation of self-trapped quantum droplets~\cite{odell00,Maucher:PRL:2011,Petrov:PRL:2015,Mazzznti:PRL:2016,Tarruell:Science:2018,Ferioli19} and the emergence of collective roton excitations \cite{Santos:PRL:2003,Henkel:PRL:2010,Tommaso:PRA:2013,Cinti:NatCom:2014,Mottl1570} that show similarities to the phenomenology of superfluid helium \cite{Landau:PR:1949}. Atomic BECs with strong dipole-dipole interactions offer an excellent platform to explore such phenomena, as demonstrated in a series of recent experimental breakthroughs~\cite{Pfau:nature:2016,Pfau:nature2:2016,Ferrier_Barbut:PRL:2016,Ferlaino:PRX:2016,Tanzi:PRL:2019,Bottcher:PRX:2019,Ferlaino:PRX:2019,Tanzi:Nature:2019,PhysRevX.11.011037,Ferlaino:NatPhys:2021,Guo:Nature:2019,Natale:PRL:2019,Tanzi:Nature:2019,Tanzi1162Science,Ferlaino:arXiv:birth}. 

These experiments~\cite{Pfau:nature:2016,Pfau:nature2:2016,Ferrier_Barbut:PRL:2016} have revealed the profound effects of quantum fluctuations~\cite{Pelster:PRA:2011,Pelster:PRA:2012}, which can stabilize the condensate~~\cite{Santos:PRA:2016,Bisset:PRA:2016,Blakie:PRA:2016,Saito:JPhysJ:2016,Santos:PRA2:2016,Blakie:PRL:2018} against collapse that would otherwise be caused by the attractive part of the dipole-dipole interaction~\cite{PhysRevLett.97.160402,Lahaye:PRL:2008}. This stabilization facilitates the formation of regular patterns or periodic density waves, triggered by the softening of roton excitations. With its broken continuous translational symmetry, the resulting state can be considered a superfluid crystal, or a supersolid state of matter, which has been long sought after since its early prediction in low-temperature  helium~\cite{Andreev:JETP:1969,Chester:PRA:1970,Leggett:PRL:1970} several decades ago. Experiments on atomic dipolar BECs have enabled detailed studies of the exotic state.
These include measurements of the supersolid excitation spectra~\cite{Guo:Nature:2019,Natale:PRL:2019}, observations of transient behaviour~\cite{Bottcher:PRX:2019,Ferlaino:PRX:2019} and the superfluid-supersolid phase-transition~\cite{Tanzi:Nature:2019,PhysRevX.11.011037,Ferlaino:NatPhys:2021}, as well as the probing of nonclassical rotational inertia~\cite{Tanzi1162Science} and temperature effects~\cite{Ferlaino:arXiv:birth}. 

While similar density-wave supersolid states can also occur in other weakly interacting quantum fluids \cite{Henkel:PRL:2010,Cinti:PRL:2010,Tommaso:PRA:2013,Cinti:NatCom:2014,odell03,PhysRevLett.108.225301,Donner:nature:2017,Ketterle:nature:2017,PhysRevA.101.043602,PhysRevX.6.021026}, dipolar BECs stand out due to the important role played by quantum fluctuations. Indeed theoretical calculations have shown that quantum fluctuations can have profound effects on the superfluid-supersolid phase transition, giving rise to a critical point in the phase diagram of condensates with two-dimensionally broken lattice symmetry and cause the emergence of different supersolid states with distinct periodic structures~\cite{Yongchang:PRL:2019}. 
While the direct identification and exploration of phase transitions is possible in theoretical calculations, condensates in experiments are intrinsically finite and therefore it is necessary to find suitable signatures to infer from the system behavior in the thermodynamic limit. Here, we address this problem and study the properties of supersolid states in trapped dipolar BECs. We seek to identify suitable parameters and trap geometries that would make the identification of distinct supersolid states possible with atom numbers that can be reached in current experiments. Surprisingly, our simulations not only show that the supersolid phases previously found in the thermodynamic limit can still be identified in remarkably small systems, but also reveal new supersolid lattice structures.

The article is organized as follows.
In the next section, we introduce the basic model employed to describe weakly interacting dipolar BECs. In section III, we outline the basic phenomenology of the thermodynamic limit and discuss the relation between particle number and density around the phase transition in order to estimate the relevant parameter regime for finite trapped BECs, considered in the ensuing section IV. This section describes emerging phases of quasi-two-dimensional condensates with in-plane harmonic  confinement. We find density patterns that arise from the finite trap geometry as well as patterns that can be connected directly to the behavior in the thermodynamic limit. Section V, focuses on typical conditions of ongoing experiments, discusses the BEC dynamics in the presence of losses and presents optimal parameters that should facilitate observations of the different supersolid states.

\section{Basic model for weakly interacting dipolar BECs}
We consider a zero-temperature quantum gas of $N$ dipolar Bosonic atoms with a mass $m$.
The positions ${\bf r}$ of the atoms are confined in all three spatial directions by a harmonic trapping potential $U({\bf r})$.
The particles interact through a short-range potential with an s-wave scattering length $a_{\rm s}$ and via long-range dipole-dipole interactions, characterized by an associated length scale $a_{\rm dd}$ \cite{Lahaye:RPP:2009}. 
The condensate wave function $\psi({\bf r})$ of the atoms is normalized to one, such that $\rho=N|\psi|^2$ defines the atomic density.

We scale spatial coordinates by $\ell=12\pi a_{\rm dd}$ and express time in units of $m\ell^2/\hbar$. The dimensionless total energy $E$ of the condensate can be written as~\cite{Pelster:PRA:2011,Pelster:PRA:2012,Santos:PRA:2016,Bisset:PRA:2016,Blakie:PRA:2016}
\begin{equation}
 \frac{E}{N}=\int \frac{|\nabla\psi|^2}{2} + U|\psi|^2+\frac{2}{5}\gamma N^{3/2}|\psi|^{5}{\rm d}{\bf r}+\frac{E_{\rm I}}{N}. \label{eq:energy}
\end{equation}
The first term corresponds to the kinetic energy of the atoms and the trapping potential $U({\bf r})$ is given by  $U(x,y,z)=\frac{1}{2}[\omega^2_{\perp}(x^2+y^2)+\omega_z^2 z^2]$. Here, $\omega_{\perp}$ and $\omega_z$ denote the dimensionless trap frequency for the in-plane and longitudinal confinement, respectively. We assume that the atoms are polarized along the $z$-direction. The third term accounts for quantum fluctuations to leading order in the strength of the atomic interactions, as given by the Lee-Huang-Yang (LHY) correction \cite{LHY1,LHY2}, where $\gamma=\frac{4}{3\pi^2}(\frac{a_{\rm s}}{3a_{\rm dd}})^{5/2}[1+\frac{3}{2}(\frac{a_{\rm dd}}{a_{\rm s}})^2]$~\cite{Pelster:PRA:2011, Pelster:PRA:2012, Blakie:PRA:2016, Santos:PRA:2016, Bisset:PRA:2016}. Note that this expression for the LHY term $\gamma$ is based on a local density approximation~\cite{Pelster:PRA:2011}, whose predictive power has been confirmed by comparisons to Quantum Monte Carlo simulations \cite{Saito:JPhysJ:2016} and experiments \cite{Tanzi:PRL:2019,Bottcher:PRX:2019,Ferlaino:PRX:2019,Tanzi:Nature:2019,Guo:Nature:2019,Natale:PRL:2019}. Despite of its simplicity, it agreed well with both Monte Carlo  simulations~\cite{Saito:JPhysJ:2016} and reasonably well with  experiments, e.g.~\cite{Tanzi:PRL:2019,Pfau:PRX:2019,Ferlaino:PRX:2019,Tilman:PRR:2019}.
The corresponding mean-field interaction energy 
\begin{equation}
E_{\rm I}=N^2\!\!\!\int\!\!\frac{a_{\rm s}}{6a_{\rm dd}}|\psi({\bf r})|^4+\frac{|\psi({\bf r})|^2}{8\pi}\!\!\int\!\! V({\bf r}-{\bf r}^\prime)|\psi({\bf r}^\prime)|^2{\rm d}{\bf r}^\prime {\rm d}{\bf r}\label{eq:Epot}
 \end{equation}
is a sum of terms describing zero-range collisions, proportional to $a_{\rm s}$, and long-range dipole-dipole interactions with the interaction potential $V({\bf r})=(1-3z^2/r^2)/r^3$. 
The equation of motion is readily found from $i\partial_t\psi=\frac{\delta E}{\delta \psi^*}$ and reads
\begin{align}\label{eq:motion}
i\partial_t\psi =&-\frac{1}{2}\Delta\psi+U\psi+\gamma N^{3/2}|\psi|^{3}\psi \nonumber\\
                &+\frac{N a_{\rm s}}{3a_{\rm dd}}|\psi|^2\psi+\frac{N}{4\pi}\int V(\mathbf{r-r^{\prime}})|\psi(\mathbf{r^{\prime}})|^2{\rm d}{\bf r}\psi.
\end{align}
The last term, that describes the dipolar mean-field interaction, can cause collapse due to the attractive part of the interaction potential $V({\bf r})$. While this effect has been long believed to prevent the formation of supersolid states~\cite{Lahaye:PRL:2008,Lahaye:RPP:2009}, it was later found that quantum fluctuations as described by the fourth term in Eq.(\ref{eq:motion}) stabilize the condensate and thereby facilitate supersolidity in dipolar condensates.

\section{Supersolid phases in the thermodynamic limit}\label{InfCase}
In order to gain some intuition about the behavior of confined systems, we begin by considering the thermodynamic limit. As we shall see below the analysis of the thermodynamic limit provides a simple relation between system parameters around the supersolid transition, which will prove useful for studying finite condensates. 

More specifically, we consider first the limit of $\omega_\perp=0$, i.e. a trapping potential $U({\bm r})=\omega^2_z z^2/2$ that confines the atoms only along the $z$-axis and therefore yields an infinitely extended quasi-2D quantum gas in the $x-y$ plane. In the superfluid phase, in which $\psi({\bf r})=\psi(z)$, we can thus define the density as $\rho=N |\psi|^2$, where $|\psi|^2\rightarrow0$ and $N\rightarrow \infty$ to yield a finite value of $\rho(z)$ and the 2D density $\rho_{\rm 2D}=\int{\rm d}z \rho(z)$.

\begin{figure}[t!]
\includegraphics[width=\columnwidth]{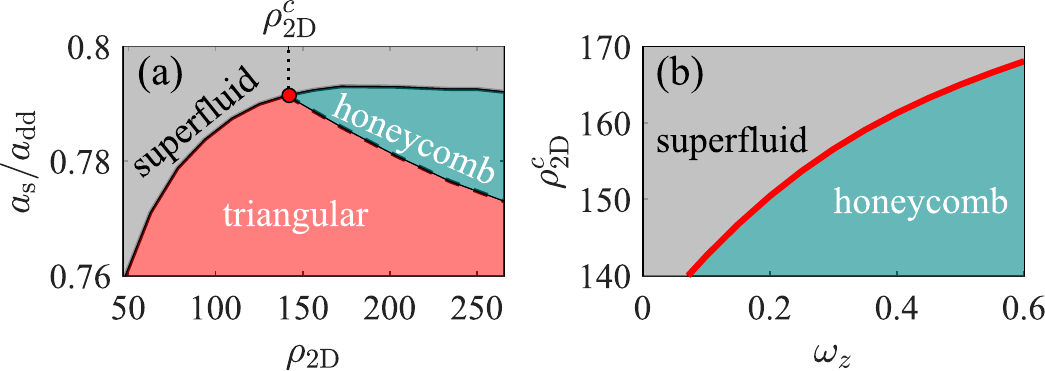}
\caption{\label{fig:second_order_line}
Phase diagram of quasi-two-dimensional dipolar BECs in the thermodynamic limit. (a) For a fixed value of $\omega_z$ one finds three different phases: a superfluid with a constant density $\rho_{\rm 2D}$, and two supersolid phases in the form of a triangular droplets crystal and a honeycomb structure. All three phases co-exist at a critical point (red) with a density $\rho_{\rm 2D}^c$, where the phase-transition is of second order. This value also yields the minimal density for which a honeycomb supersolid is possible. The dependence of this critical density on the longitudinal trap frequency is shown in panel (b).}
\end{figure}

The competition between the dipolar and short-range interaction gives rise to a phase transition between the superfluid state with a constant in-plane density and a spatially modulated state corresponding to a density-wave supersolid \cite{PhysRevLett.115.075303,PhysRevA.99.041601,Bottcher:PRX:2019,Ferlaino:PRX:2019,Tanzi:PRL:2019}. As shown in Fig.~\ref{fig:second_order_line}(a), the underlying phase diagram can be more complex and features two competing types of supersolids in the form of a triangular droplet lattice or a regular honeycomb density pattern~\cite{Yongchang:PRL:2019}. 
All three phases coexist at a critical point at which the supersolid phase transition is of second order type and hence a continuously growing density wave upon varying the atomic density or interactions takes place.
One can approximately analyze this behavior within the following variational procedure. We consider the energy difference between the unmodulated ground state and a state containing small periodic density modulations of the form~\cite{Yongchang:PRL:2019}
\begin{equation}
    \rho({\bf r})=\rho_0(z)\left( 1 + A \sum_{j=1}^3  \cos ({\bm k}_{j}{\bf r}) \right).
    \label{eq:pert_TF}
\end{equation}
where we approximate the $\rho_0(z) = \frac{3\rho_{\rm 2D}}{4\sigma_z}\left(1-\frac{z^2}{\sigma^2_z}\right)$ of the unmodulated superfluid state by a simple Thomas-Fermi profile, with $\sigma_z$ is given by $\sigma_z = \left(\frac{\rho_{\rm  2D}(a_{\rm s}/a_{\rm dd}+2)}{2\omega_z^2}\right)^{1/3}$. The variational parameters $A$ and ${\bm k}_j$ correspond to the amplitude and the wave vector of the periodic modulation in the $x-y$ plane. Optimizing the energy with respect to the angles between the wave vectors shows that triangular modulations with $\sum_{j=1}^3 {\bf k}_j=0$ and $|{\bm k}_j|=k$ present the most favourable configuration. Substituting Eq.~(\ref{eq:pert_TF}) into Eq.~(\ref{eq:energy}) and expanding the result in $|A|\ll1$ yields 
\begin{equation}\label{eq:lin_pert}
\frac{\Delta E}{N}=a^{(2)}A^2+a^{(3)}A^3+a^{(4)}A^4
\end{equation}
for the energy difference between the modulated and unmodulated state, where
\begin{eqnarray}
a^{(2)}&=&\frac{3}{4}\bigg[ \frac{k^2}{4} +\frac{ a_{\rm s} \rho_{\rm 2D}}{5 a_{\rm dd} \sigma_z} +\frac{45\pi \gamma}{128}\left(\frac{3\rho_{\rm 2D}}{4\sigma_z}\right)^{3/2} \\ \nonumber
&~&+\frac{3 \rho_{\rm 2D}}{4\sigma_z} \bigg(f(k \sigma_z) -\frac{4}{15} \bigg)\bigg],\\
a^{(3)}&=&-\frac{3}{32}\left[k^2 - \frac{15\pi \gamma}{32}\left(\frac{3\rho_{\rm 2D}}{4\sigma_z}\right)^{3/2} \right],\\
a^{(4)}&=&\frac{15}{64}\left[ k^2 -\frac{45\pi \gamma}{512}\left(\frac{3\rho_{\rm 2D}}{4\sigma_z}\right)^{3/2}\right],
\label{eq:coeff}
\end{eqnarray} with $f(x)=({3-3x^2+2x^3-3(1+x)^2 e^{-2x}})/{x^5}$. The roots of Eq.~(\ref{eq:lin_pert}) yield the location of the phase transition which is determined only by the ratio $a_{\rm s}/a_{\rm dd}$, the 2D density $\rho_{\rm 2D}$ and the longitudinal width of the cloud $\sigma_z$, which can be controlled by the trapping frequency $\omega_z$. This approach yields an accurate description of the phase diagram in Fig.~\ref{fig:second_order_line} around the critical point where $|A|\ll1$, whereby the triangular lattice corresponds to $A>0$, while negative amplitudes $A<0$ yield the honeycomb structure. As shown Fig.~\ref{fig:second_order_line}(a), the latter requires to cross a density $\rho_{\rm 2D}^{c}$ set by the critical point, which therefore defines the minimum density necessary to observe the honeycomb state. Its dependence on the longitudinal trap frequency is depicted in~\reffig{fig:second_order_line}(b) and shows that weaker trapping facilitates supersolidity at lower atomic densities $\rho_{\rm 2D}$.

\begin{figure}[t!]
\centering
\includegraphics[width=0.9\columnwidth]{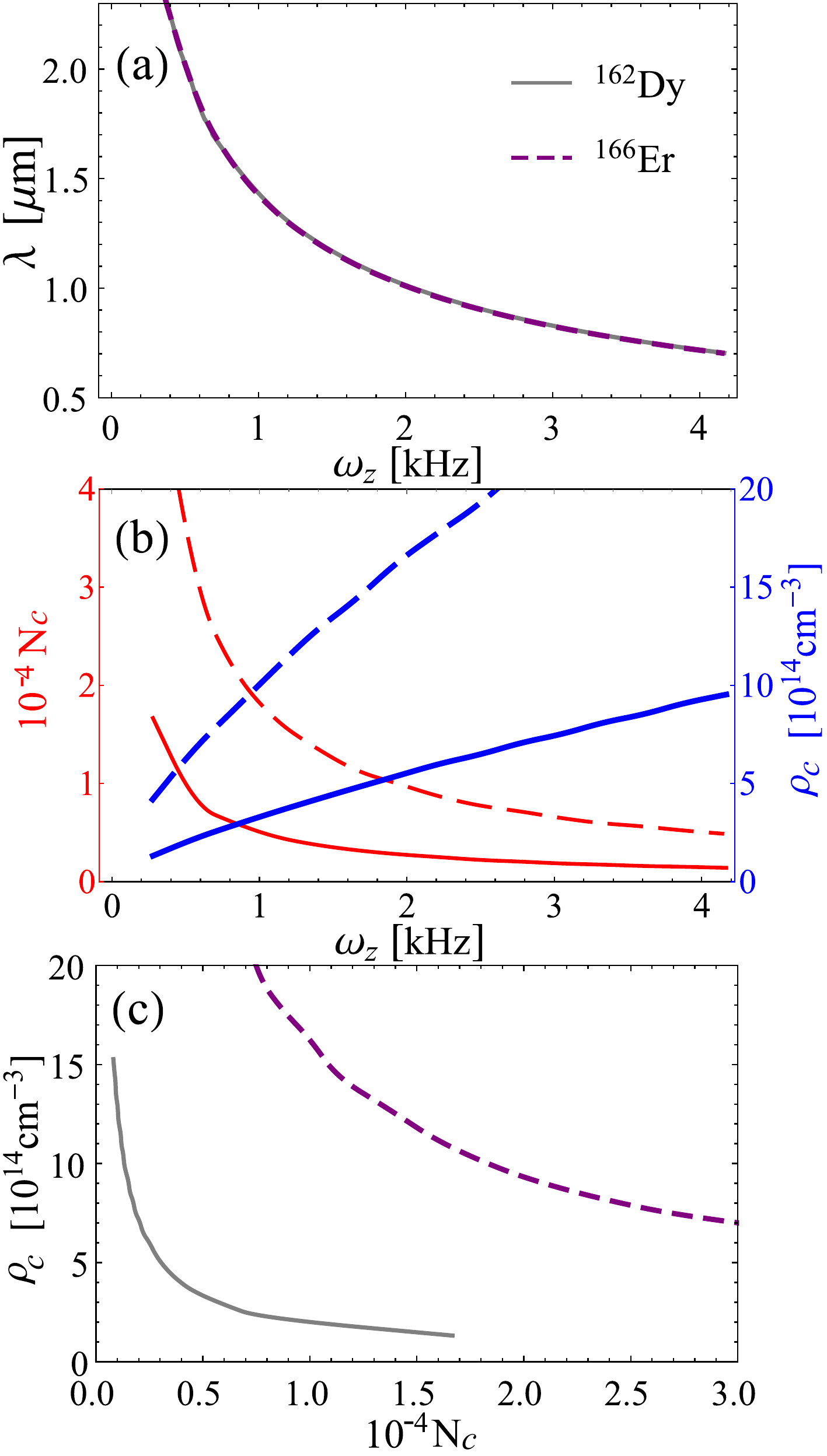}
\caption{
Panel (a) shows the wavelength $\lambda$ of the supersolid density-wave state as a function of the trap frequency, while panel (b) presents the $\omega_z$-dependence of the critical density $\rho_c=3\rho_{\rm 2D}^c/(4\sigma_z)$ and particle number $N_c=2\rho_{\rm 2D}^{c}\lambda^2/\sqrt{3}$ per unit cell of the honeycomb lattice. In (c) we show the resulting relation between $\rho_c$ and $N_c$, showing that reaching the critical point at a lower density requires an increases of the particle number per unit cell.}
\label{N_rho}
\end{figure}

On the other hand, weaker trapping corresponds to a larger width $\sigma_z$ and therefore implies a larger particle number for a given 2D density $\rho_{\rm 2D}$. While the strength of the trapping potential can be varied broadly in experiments, reaching large particle numbers presents one of the main experimental challenges. We therefore seek to minimize the particle number $N_c=\rho_{\rm 2D}^{c}V_{\rm unit}$ per unit cell. Here $V_{\rm unit}=2\lambda^2/\sqrt{3}$ is the unit cell area, where the length 
$\lambda=\frac{2\pi}{k}$ is given by the wave vector of the modulated density Eq.~(\ref{eq:pert_TF}) at the critical point.

Figure \ref{N_rho} shows the obtained relation between the most relevant parameters at the critical point. The results are shown for dipolar condensates of Erbium and Dysprosium atoms, which have comparably strong dipole-dipole interactions with $a_{\rm dd}=65.5a_0$ \cite{PhysRevX.6.041039} and $a_{\rm dd}=132a_0$ \cite{Pfau:nature:2016}, respectively, and have been used in a series of recent experiments. As shown in Fig.~\ref{N_rho}(a), a stronger longitudinal confinement yields smaller length scales for the transverse density-wave pattern in the supersolid phase, which is independent of the dipolar interaction strength $a_{\rm dd}$ and therefore identical for both atomic species. The ability to control and to reduce the length scale of the emerging lattice is crucial since the total size of the BEC is typically limited in experiments. 

Below, we will use these insights in order to optimize the parameters and minimize the required atom numbers for confined condensates under realistic experimental conditions. 

\begin{figure}[t!]
\includegraphics[width=\columnwidth]{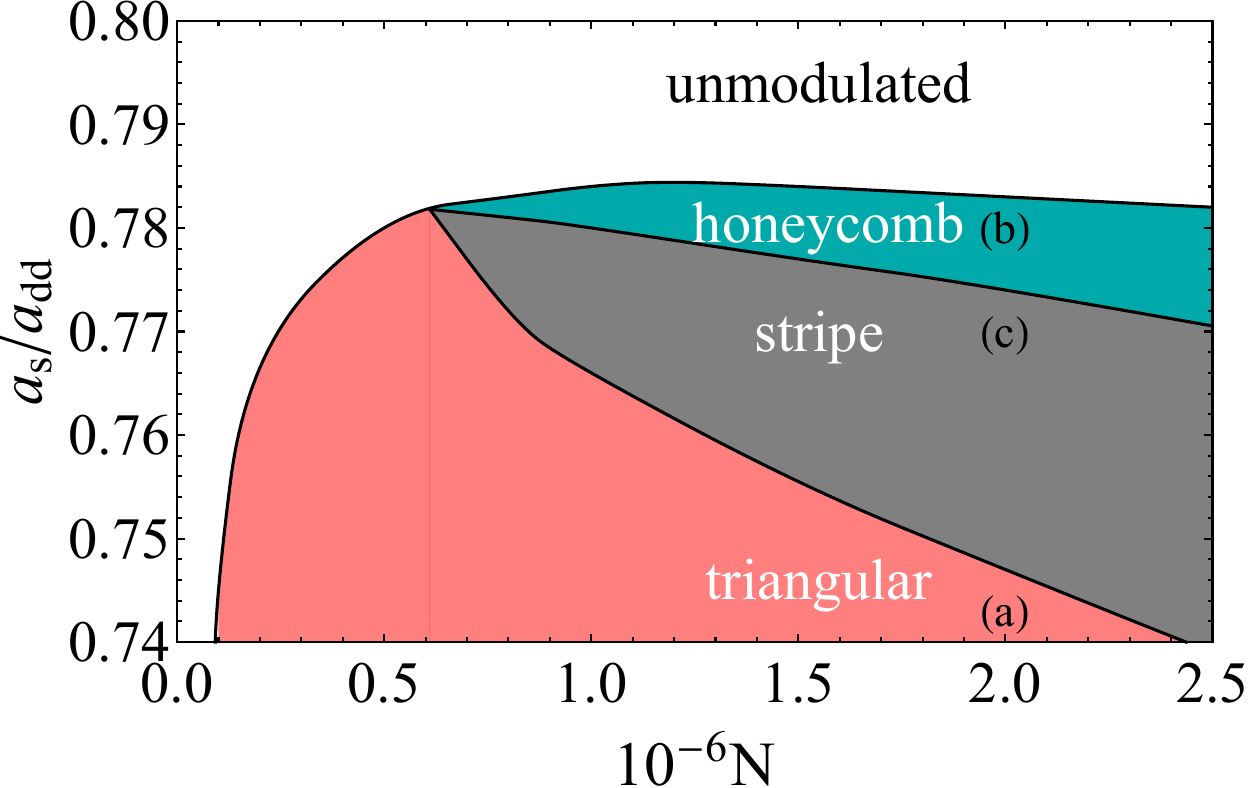}
\caption{\label{PD}Phase diagram of finite condensates in a weak trap with $\omega_\perp=167\ {\rm Hz}$ and $\omega_z=444\ {\rm Hz}$ as function of $a_{\rm s}/a_{\rm dd}$ and the total number $N$ of particles. 
Note the qualitative similarities with the thermodynamic limit shown in Fig.~\ref{fig:second_order_line}~\cite{Yongchang:PRL:2019}, including the critical point at which the triangular, honeycomb and unmodulated superfluid coexist. In addition, however, one finds new supersolid states composed of 
striped density waves. Exemplary density profiles of the states are shown in Fig.~\ref{GroundStates}. 
}
\end{figure}

\section{Finite Systems}
As a preliminary step, we now consider the behavior of finite but large systems with $\sim10^6$ atoms, in order to gain more intuition about the effect of the trapping potential on the density patterns in the thermodynamic limit, as discussed above. Supersolid states in finite, trapped BECs have been studied theoretically in Refs.~\cite{PhysRevLett.115.075303} and in very recent experiments~\cite{Ferlaino:arXiv:2021}.

To find the various phases numerically we seed different initial conditions and employ imaginary-time evolution and Fourier split-step methods on a uniformly spaced grid ($\Delta l=0.52{\rm \mu m}$, $\Delta t=0.07{\rm \mu s}$) to relax the system to the respective local energy minimum. 
To describe the dipolar interaction accurately we use the cut-off described in \cite{Ronen:PRA:2006} and ensure that the extent of the numerical box corresponds to roughly three times the localised Bose-Einstein condensate. By considering the energy crossings between the converged states we can identify the boundary between different phases. 

\begin{figure}[t!]
\centering
   \includegraphics[width=\columnwidth]{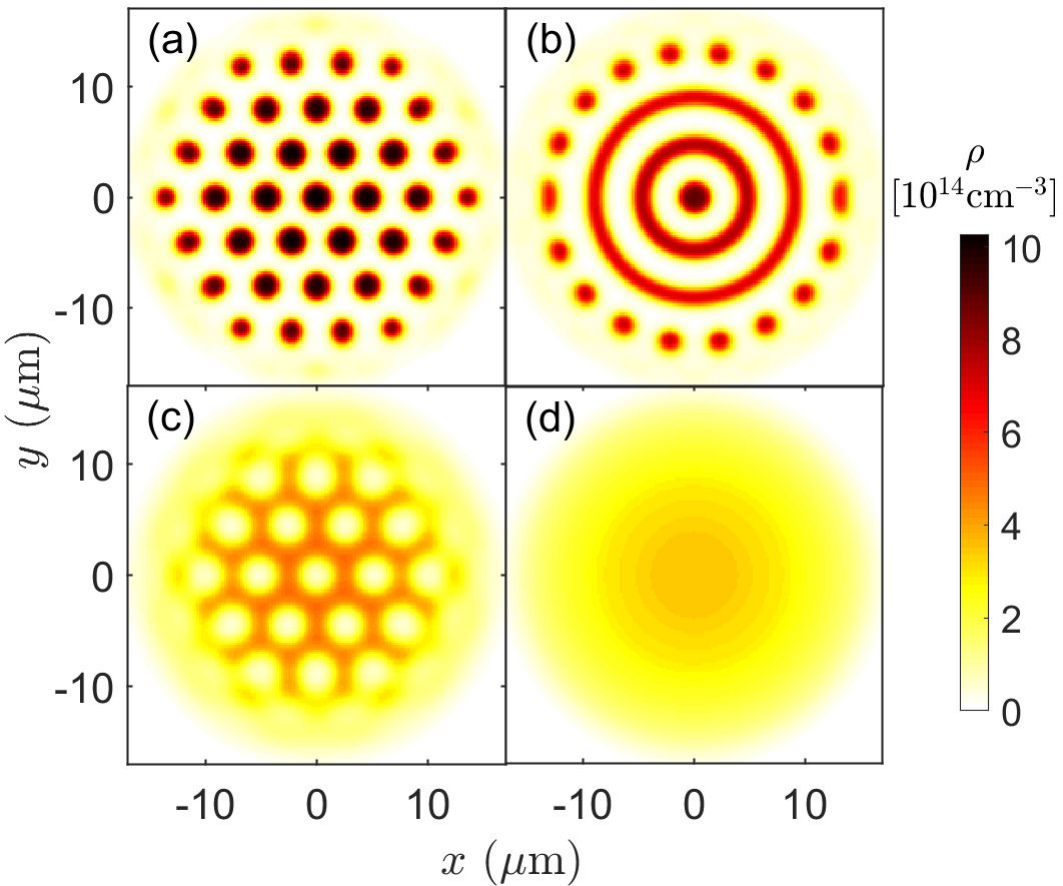}
\caption{Ground state density profiles for $N=2\times 10^6$. One finds (a) a triangular droplet lattice for $a_{\rm s}/a_{\rm dd}=0.74$, (b) a honeycomb structure for $a_{\rm s}/a_{\rm dd}=0.78$, 
(c) a stripe phase for $a_{\rm s}/a_{\rm dd}=0.77$ and (d) a metastable ring state for the same value $a_{\rm s}/a_{\rm dd}=0.77$.
The remaining parameters are the same as in Fig.~\ref{PD}. 
\label{GroundStates}}
\end{figure}

A ground state phase diagram that illustrates the role of the particle number is shown in Fig.~\ref{PD}. While the depicted transition lines do not represent phase transitions in a thermodynamic sense and depend on the in-plane trapping geometry, one can nevertheless define sharp transition lines by comparing the energies of different low-lying stationary states of the trapped condensate. While much of the depicted particle numbers lie well above achievable values of current experiments, the corresponding densities involved range from $10^{14}\ {\rm cm^{-3}}$ to $10^{15}\ {\rm cm^{-3}}$, which corresponds to the typical experimental values~\cite{Pfau:nature:2016,Pfau:nature2:2016,Ferlaino:PRX:2019,Tanzi:PRL:2019}. From the density profiles of the corresponding states (Fig.~\ref{GroundStates}), one can clearly identify the unmodulated superfluid state, the triangular lattice [Fig.~\ref{GroundStates}(a)] and the honeycomb state [Fig.~\ref{GroundStates}(b)], discussed above for the thermodynamic limit. 
In addition, we find an extended parameter region in which the BEC develops a region in which striped density waves emerge as the groundstate [Fig.~\ref{GroundStates}(c)]. Similar striped states have also been found in quantum Monte-Carlo simulations~\cite{Boninseni:LowT:2019}. Furthermore, radial lattices of concentric rings [Fig.~\ref{GroundStates}(d)] appear as metastable states reflecting the radial symmetry of the trap. Radial and stripe phases are often energetically extremely close  and almost degenerate. 
Interestingly, these states only emerge beyond a critical atom number in between the parameter region of the triangular lattice and the honeycomb state (see Fig.~\ref{PD}). Consequently, the trapped condensate resembles the states found in the thermodynamic limit for both small and large values of $a_{\rm s}/a_{\rm dd}$. Remarkably, such regular structures can already be identified clearly for comparably small system sizes that contain only a few unit cells of the formed lattices.

Yet, the particle numbers are still considerably larger than what is achievable in current experiments \cite{Guo:Nature:2019,Tanzi:Nature:2019,Pfau:PRX:2019,Ferlaino:PRX:2019,Ferlaino:arXiv:2021}. We note, however, that the general structure of the underlying phase diagram in Fig.~\ref{PD} closely resembles that of the thermodynamic limit [Fig.~\ref{fig:second_order_line}(a)], such that we use the insights from the preceding section to optimize the parameters around the critical point where the different supersolid phases coexist. 

\section{Small particle numbers}
\begin{figure}[t!]
\centering
  \includegraphics[width=\columnwidth]{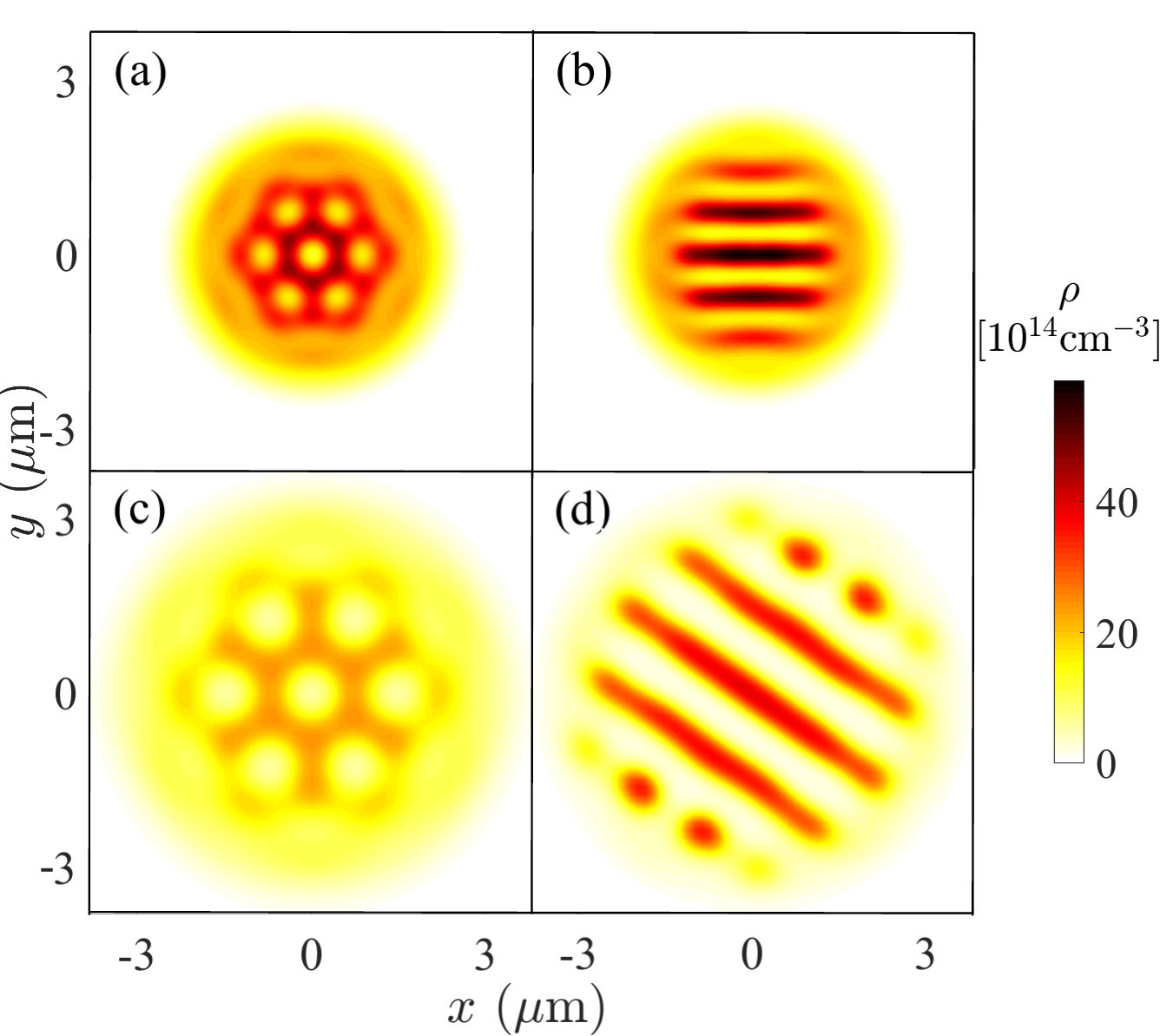}
\caption{
Groundstate density profiles for (a , b) $N=4\times 10^4$  Dysprosium atoms with $\omega_\perp=4{\rm kHz}$ and $\omega_z=16{\rm kHz}$ and (c, d) $N=10^5$ Dysprosium atoms with $\omega_\perp=1.8{\rm kHz}$ and $\omega_z=5.3{\rm kHz}$). We used $a_{\rm s}/a_{\rm dd}=0.48$ in (a), $a_{\rm s}/a_{\rm dd}=0.476$ in (b), $a_{\rm s}/a_{\rm dd}=0.59$ in (c), and $a_{\rm s}/a_{\rm dd}=0.57$ in (d). 
}
\label{Honeycomb}
\end{figure}

Following the preceding discussions, supersolidity at smaller particle numbers can be achieved by tightening the longitudinal confinement. 
This is demonstrated in Figs.~\ref{Honeycomb}(a,c), which show the honeycomb state for atom numbers of $4\times 10^4$ and $10^5$, whereby the honeycomb symmetry can be clearly identified despite the small overall size of the system. 
As in section IV we recover that a decreasing interaction strength leads to a further breaking of the rotational symmetry and the emergence of a striped phase, which neither reflects the rotation symmetry of the trapping potential nor the triangular symmetry of the supersolid ground states in the thermodynamic limit. Similar striped supersolid phases have been predicted and observed in BECs with optically induced spin-orbit coupling \cite{PhysRevLett.105.160403,PhysRevLett.107.150403,PhysRevLett.108.225301,Li2017Nature,PhysRevA.102.053308} as well as optical feedback-systems~\cite{baio2021multiple}. Here they emerge without additional symmetry-breaking terms that would determine the orientation of the formed stripes. 

\begin{figure}[t!]
\centering
  \includegraphics[width=0.9\columnwidth]{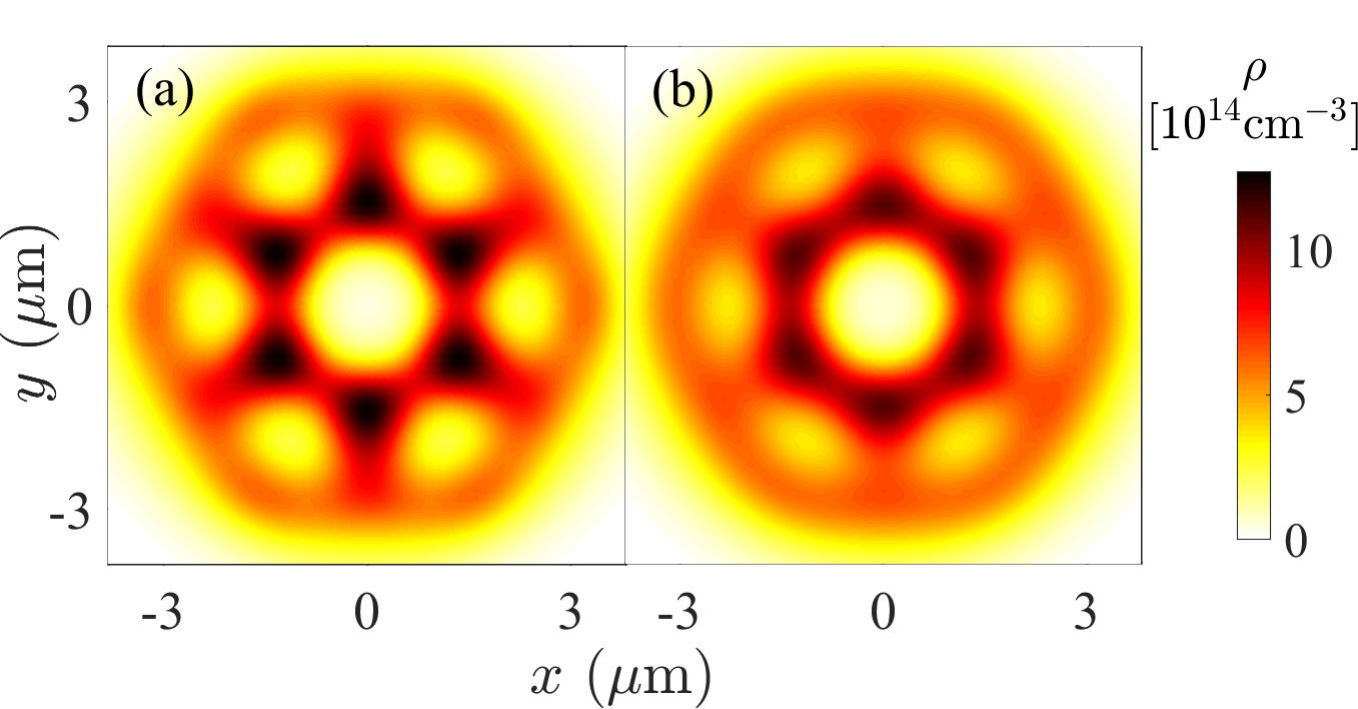}
\caption{
Ground state density profiles of a Dysprosium condensate confined in the trapping potential Eq.~(\ref{eq:box}) with $\omega_\perp=0.89{\rm kHz}$,  $\omega_z=1.9{\rm kHz}$, $R=0.53{\rm \mu m}$, and $N=10^5$. The interaction strengths are (a) $a_{\rm s}/a_{\rm dd}=0.469$ and (b) $a_{\rm s}/a_{\rm dd}=0.472$. 
}
\label{BoxPotential}
\end{figure}

The required increase of the trapping frequency to lower the required particle number, however, entails an increase of the atomic density. The associated growth of three-body losses will eventually limit the experimental observability of these states under the conditions discussed above. As we shall see below, different trap geometries make it possible to alleviate this limitation. 
To be specific, we consider a simple potential of the form
\begin{equation}\label{eq:box}
V_{\rm box} = \left\{ \begin{array}{ll}
\frac{1}{2}\omega_z^2 z^2 & \textrm{if $r_\perp\leq R$}\\
\frac{1}{2}\left[\omega_\perp^2 (r_\perp-R)^2 + \omega_z^2 z^2\right] & \textrm{if $r_\perp>R$},
\end{array} \right.
\end{equation}
which features a flat core of radius $R$, beyond which the in-plane potential increases quadratically with the radial distance $r_\perp=\sqrt{x^2+y^2}$. As shown in Fig.~\ref{BoxPotential} this potential can indeed promote supersolid ground state with the discussed honeycomb structures for $10^5$ Dysprosium atoms at a significantly reduced density. 

\begin{figure}[b!]
\centering
\includegraphics[width=0.8\columnwidth]{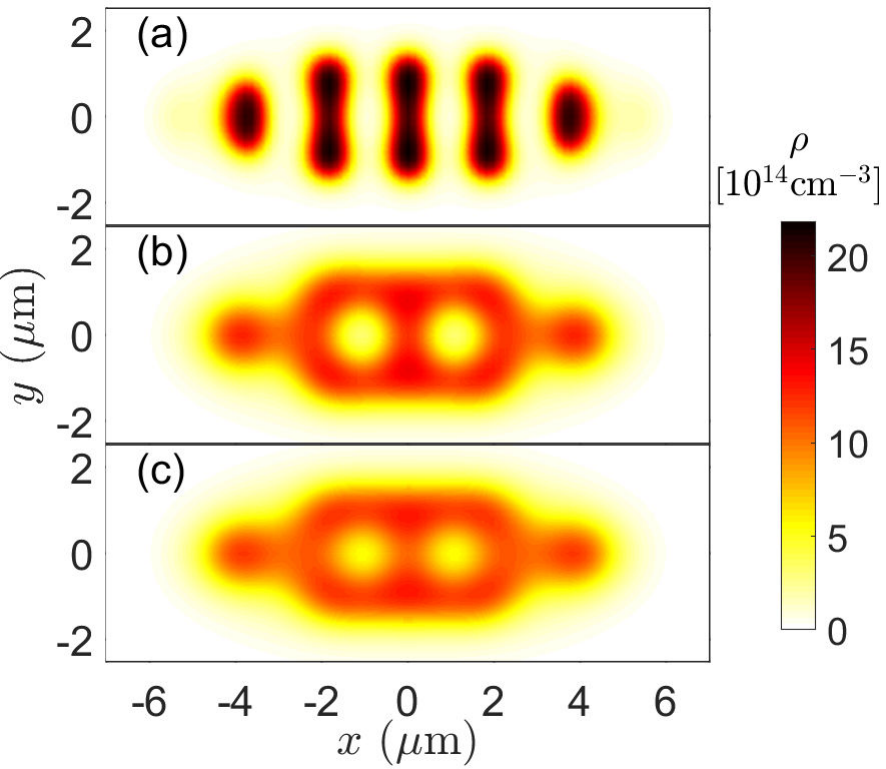}
\caption{Ground state density profiles of a Dysprosium condensate in a harmonic trap with $(\omega_x,\ \omega_y,\ \omega_z)=(0.63,\ 1.24,\ 1.67){\rm kHz}$ and $N=10^5$. The different depicted states have been obtained for (a) $a_{\rm s}/a_{\rm dd}=0.654$, (b) $a_{\rm s}/a_{\rm dd}=0.672$, and (c) $a_{\rm s}/a_{\rm dd}=0.674$.}
\label{CigarTrap}
\end{figure}

\begin{figure*}[t!]
    \centering
    \includegraphics[width=0.82\textwidth]{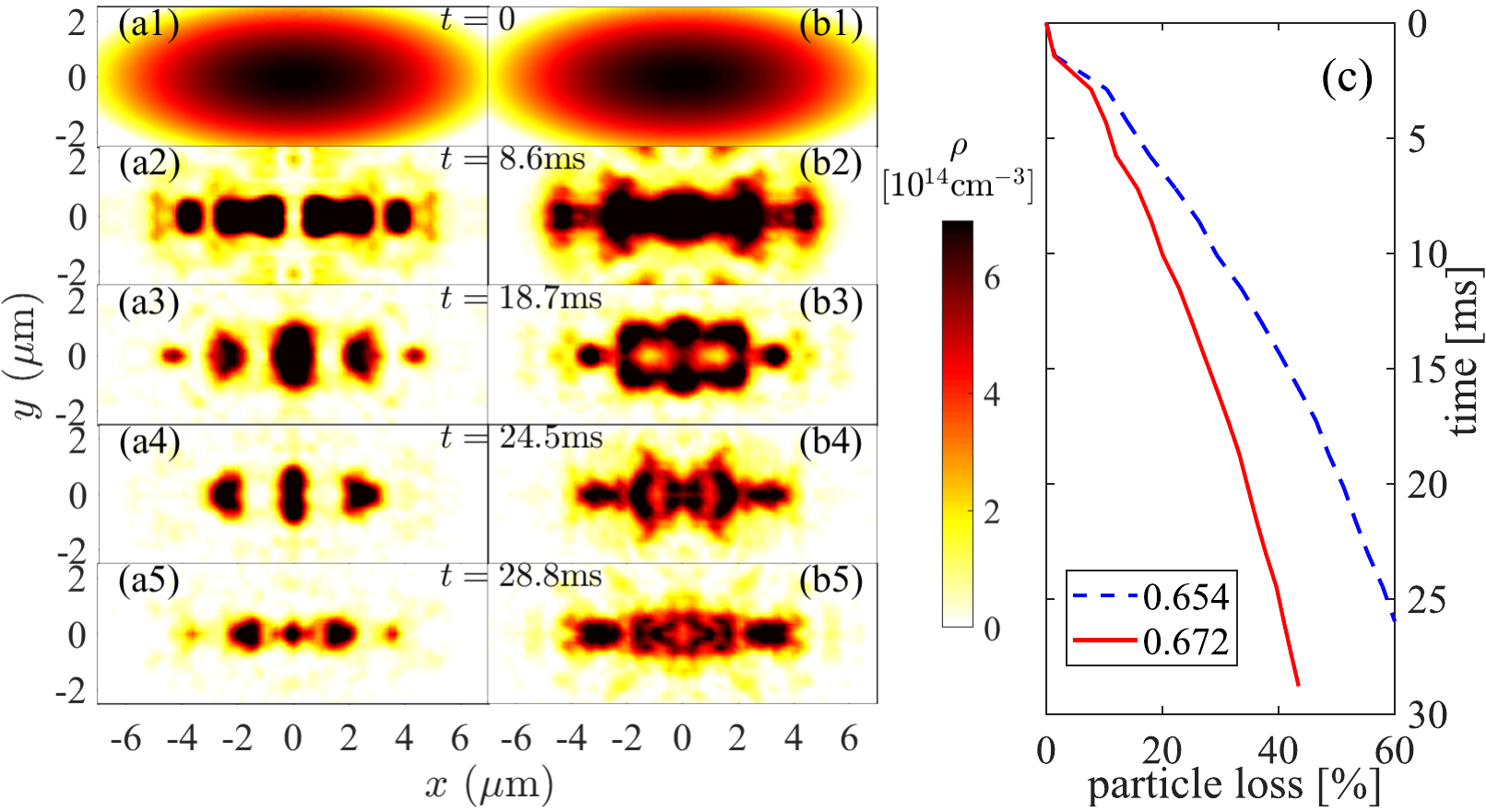}
    \caption{BEC dynamics following an interaction quench for different values of $a_{\rm s}/a_{\rm dd}$. The initial state in panels (a1)-(b1) shows the Thomas-Fermi density of the unmodulated ground state for $a_{\rm s}/a_{\rm dd}=0.7$ and $10^5$ Dysprosium atoms in a harmonic trap with $(\omega_x,\ \omega_y,\ \omega_z)=(0.63,\ 1.24,\ 1.67){\rm kHz}$, as in Fig.~\ref{CigarTrap}.  The remaining panels (a)-(b) show the density at the indicated times following an interaction quench at $t=0$ to  (a) $a_{\rm s}/a_{\rm dd}=0.654$, and (b) $a_{\rm s}/a_{\rm dd}=0.672$. The simulations account for three-body losses with a loss-rate of $L_3=1.5 \times 10^{-41} {\rm m^6/s}$ \cite{Tilman:PRR:2019}, and the resulting decrease of the atom number is shown in panel (c).}
    \label{Quench}
\end{figure*}

One may further reduce the size of the system by departing from the rotational trap geometry, as done in recent experiments \cite{Tanzi:PRL:2019,Guo:Nature:2019,Tanzi:Nature:2019,Pfau:PRX:2019,Ferlaino:PRX:2019,Ferlaino:arXiv:2021}. 
Fig.~\ref{CigarTrap} shows an example for a harmonically confined BEC with three different trapping frequencies, $(\omega_x,\ \omega_y,\ \omega_z)=(0.63,\ 1.24,\ 1.67){\rm kHz}$ for $N=10^5$ particles. For the smallest shown value of $a_{\rm s}/a_{\rm dd}$, the ground state resembles the striped phase found for the radially symmetric confinement discussed above [Fig.~\ref{CigarTrap}(a)], and undergoes a clear structural transition as $a_{\rm s}/a_{\rm dd}$ is increased [see Figs.~\ref{CigarTrap}(b) and (c)]. While the small system size does not permit to directly associate the formed states with the previously found lattice states, the clear density minimum surrounded by an edge of maximum density is reminiscent of the honeycomb structure found above for larger systems.

In order to explicitly address the density limitation and determine the associated effect of particle loss, we finally discuss real-time simulations of the condensate dynamics, taking into account three-body losses with the known loss rate of $L_3=1.5 \times 10^{-41} {\rm m^6/s}$~\cite{Tilman:PRR:2019} for Dysprosium atoms. We consider the same trap geometry as in Fig.~\ref{CigarTrap} and initialize the simulation with a Thomas-Fermi profile
\begin{equation}
\psi(t=0)=\rho_0^{1/2} \sqrt{1-x^2/\sigma_x^2-y^2/\sigma_y^2-z^2/\sigma_z^2}
\end{equation}
with $\rho_0=7\times 10^{14}{\rm cm^{-3}}$, $\sigma_x=7.9{\rm \mu m}$, $\sigma_y=3.2{\rm \mu m}$, and $\sigma_z=3.4 {\rm \mu m}$ to approximates the actual ground state of the BEC for a scattering length of $a_{\rm s}/a_{\rm dd}=0.7$. The density profile of this initial state in shown in  Figs.~\ref{Quench}(a1,b1). Subsequently, we quench the scattering length to a smaller value and simulate the ensuing condensate dynamics in the presence of three-body losses. 

Figures~\ref{Quench}(a1)-(a5) show the obtained time evolution for a scattering length of $a_{\rm s}/a_{\rm dd}=0.654$, for which the ground state resembles a striped supersolid [see Fig.~\ref{CigarTrap}(a)]. Indeed, the condensates forms clear regular density maxima as observed in previous experiment and similar to the ground state shown in Fig.~\ref{CigarTrap}(a). More importantly though, the condensate dynamics is markedly different for larger values of $a_{\rm s}/a_{\rm dd}$, and clearly develops a density profile that closely resembles the ground state shown in Figs.~\ref{CigarTrap}(b) and (c) for the same scattering lengths. While this pattern only emerges for a finite time before it disintegrates due to three-body losses the associated timescale of a few milliseconds is sufficiently long to be resolved in current experiments \cite{Tanzi:PRL:2019,PhysRevX.11.011037,Ferlaino:arXiv:birth}. 

\section{conclusions}
In summary, we have shown that the interplay of short- and long-range interactions with external confinement leads to a rich spectrum of density-wave supersolid phases in dipolar BECs. By exploring the broad parameter space of interactions strengths, particle numbers, and trapping geometries, we have identified four supersolid states with distinctively broken spatial symmetries, that reflect the triangular-lattice symmetry of the thermodynamic limit and a stripe phase which does not feature a triangular symmetry nor the symmetry of the trap. 
We found furthermore metastable states reflecting the rotational trap symmetry in the form of concentric rings. 
We have focused here on relatively small systems to investigate the observability of these states within current experimental capabilities. 
The very recent further investigations on the appearance of glassy labyrinth-like states~\cite{hertkorn2021pattern} underpin the richness of physics in this system. 


We hope that the results of this work can guide future experimental studies of the rich phenomenology of supersolidity in dipolar BECs, this includes the effects of thermal fluctuations~\cite{Ferlaino:NatPhys:2021,Ferlaino:arXiv:birth} on the different predicted lattice structures.

\section{Acknowledgement}

We are grateful to Jens Hertkorn, Tilman Pfau and his group at the University of Stuttgart and Francesca Ferlaino for fruitful discussions and helpful comments. 
This work was supported by the DFG through the SPP1929, and by the Danish National Research Foundation through the Center of Excellence ``CCQ" (Grant agreement no.: DNRF156).

\end{document}